\pdfoutput=1
\PassOptionsToPackage{pdftex,
pdfencoding=auto,
pdfnewwindow=true,
pdfusetitle=true,
bookmarks=true,
bookmarksnumbered=true,
bookmarksopen=false,
pdfpagemode=UseThumbs,
bookmarksopenlevel=1,
pdfpagelabels=true,
breaklinks=true,
colorlinks=true,
unicode=true,
pdfborder={0 0 0},
backref=false
}{hyperref}
\PassOptionsToPackage{usenames,dvipsnames,pdftex}{xcolor}
\documentclass[superscriptaddress,aps,pra,nofootinbib,twocolumn,twoside,reprint,letterpaper,longbibliography,showpacs,10pt,floatfix]{revtex4-2}
\usepackage[utf8]{inputenx}
\usepackage{amsmath,bbm}
\usepackage{amssymb}
\usepackage{graphicx}
\usepackage{color}
\usepackage{amsthm}
\usepackage{dsfont}
\usepackage{float}
\usepackage{cancel}
\usepackage{booktabs}
\usepackage[alwaysadjust,flushleft,pointlessenum]{paralist}
\setdefaultenum{\bfseries1.}{}{}{}

\usepackage[colorlinks=true, allcolors=blue]{hyperref}
\usepackage{amsthm}
\usepackage{amsmath}
\usepackage{amssymb}
\usepackage{graphicx}
\usepackage{color}
\usepackage{enumitem}   

\newcommand{\W}{\mathrm{\W}}
\newcommand{\GHZ}{\mathrm{\GHZ}}

\newcommand{\beq}{\begin{equation}}
\newcommand{\eeq}{\end{equation}}
\newcommand{\beqa}{\begin{eqnarray}}
\newcommand{\eeqa}{\end{eqnarray}}

\newcommand{\ket}[1]{\ensuremath{\left|#1\right\rangle}}

\renewcommand{\today}{\number\day\space\ifcase\month\or
   January\or February\or March\or April\or May\or June\or
   July\or August\or September\or October\or November\or December\fi
   \space\number\year}

\definecolor{myurlcolor}{rgb}{0,0,0.7}
\definecolor{myrefcolor}{rgb}{0.8,0,0}
\definecolor{purple}{RGB}{128,0,128}
\definecolor{ultramarine}{RGB}{63, 0, 255}
\definecolor{medblue}{RGB}{0, 0, 100}
\definecolor{googleblue}{RGB}{34, 0, 204}
\definecolor{panblue}{RGB}{0,24,150}
\definecolor{carmine}{RGB}{150, 0, 24}
\definecolor{gray}{RGB}{150, 150, 150}

\newcommand{\term}[1]{\textcolor{medblue}{\textbf{#1}}}
\usepackage{amsmath,amsfonts,amssymb,amsthm,mathtools,bm,graphicx,verbatim,mathrsfs,units}
\usepackage{cprotect} %

\newtheorem{thm}{Theorem}

\newtheorem{prop}[thm]{Proposition}

\newtheorem{definition}[thm]{Definition}

\newtheoremstyle{defblock}{0.7\topsep}{0pt}{}{}{}{: }{0pt plus 1pt minus 1pt}{\thmname{\bfseries{#1}}\thmnumber{\bfseries{#2}}\color{medblue}\bfseries\thmnote{#3}}
\theoremstyle{defblock}

\theoremstyle{remark}

\newcommand{\same}[0]{{\textsf{Same}}}
\newcommand{\Bell}[0]{{\textsf{Bell}}}

\newcommand{\tagprop}[1]{\tag{\hyperref[#1]{P\ref{#1}}}}

\usepackage{microtype}
\microtypecontext{spacing=nonfrench}
\microtypesetup{
expansion={true,nocompatibility},
protrusion={true,nocompatibility},
activate={true,nocompatibility},
tracking=true,
kerning=true,
spacing={true}
}
\usepackage[all=normal,floats=tight,mathspacing=tight,wordspacing=tight,paragraphs=normal,tracking=tight,charwidths=tight,mathdisplays=normal,sections=normal,margins=normal]{savetrees}
\everypar=\expandafter{\the\everypar\loosness=-1 }
\usepackage[style=american,autopunct=true]{csquotes}

\usepackage[scr=boondoxupr]{mathalfa} %

\usepackage{hyperref}
\hypersetup{
linkcolor=carmine,
citecolor=googleblue,
urlcolor=panblue,
anchorcolor=OliveGreen}

\AtBeginDocument{%
    \newwrite\bibnotes
    \def\bibnotesext{Notes.bib}
    \immediate\openout\bibnotes=\jobname\bibnotesext
    \immediate\write\bibnotes{@CONTROL{REVTEX42Control}}
    \immediate\write\bibnotes{@CONTROL{%
    apsrev42Control,editor="0",pages="0",title="0",year="1"}}
     \if@filesw
     \immediate\write\@auxout{\string\citation{apsrev42Control}}%
    \fi
}

\usepackage{placeins}%

\begin{document}

\begin{abstract}

We show that some tripartite quantum correlations are inexplicable by any causal theory involving bipartite nonclassical common causes and unlimited shared randomness.
This constitutes a device-independent proof that \emph{Nature's nonlocality is fundamentally at least tripartite} in every conceivable physical theory --- no matter how exotic.
To formalize this claim we are compelled to substitute Svetlichny's historical definition of genuine tripartite nonlocality with a novel theory-agnostic definition tied to the framework of Local Operations and Shared Randomness (LOSR).
A companion article [PRA. 104, 052207 (2021)] generalizes these concepts to any $N\geq3$ number of parties, providing experimentally amenable device-independent inequality constraints along with quantum correlations violating them, thereby certifying that Nature's nonlocality must be \emph{boundlessly} multipartite.
\end{abstract}

\title{No Bipartite-Nonlocal Causal Theory Can Explain Nature's Correlations}

\date{\today}

\author{Xavier Coiteux-Roy}
\email{xavier.coiteux.roy@usi.ch}
\affiliation{Faculty of Informatics, Università della Svizzera italiana, Lugano, Switzerland.}

\author{Elie Wolfe}
\email{ewolfe@perimeterinstitute.ca}
\affiliation{Perimeter Institute for Theoretical Physics, Waterloo, Ontario, Canada.}

\author{Marc-Olivier Renou}
\email{Marc-Olivier.Renou@icfo.eu}
\affiliation{ICFO-Institut de Ciencies Fotoniques, The Barcelona Institute of Science and Technology, Castelldefels (Barcelona), Spain.}

\maketitle

\emph{Introduction.---}
Nonlocality is one of the most common-sense challenging, but nevertheless well-established, properties of quantum physics~\citep{EPR,bell1964einstein}. 
Two or more parties measuring a shared entangled quantum state can obtain correlated outputs which resist explanation in terms of any local hidden variable model.
Understanding of the concept of nonlocality and of its manifestations has captivated the attention of hundreds of researchers spanning decades, see Ref.~\cite{Brunner2014} and references therein. Seminal milestones include the development of tasks inaccessible with only classical resources such as the CHSH game~\cite{CHSHOriginal}, celebrated experimental demonstrations~\cite{Freedman1972,Aspect1981,Tittel1998,Hensen2015,Shalm2015,Giustina2015,Rosenfeld2017}, and the device-independent certification of experimental apparatuses taken as black boxes~\cite{Mayers1998,Acin2006,Acin2007,Pironio2010,Arnon2016}. 

The bipartite scenario is arguably the most studied. However, scenarios with more that two parties exhibit certain valuable features which are qualitatively distinct from those of the bipartite scenario. For instance, tripartite quantum scenarios can demonstrate a stronger version of Bell's theorem~\cite{Greenberger1990GHZ}. 
More generally, the nonlocality of multipartite chains of bipartite Bell inequalities decays to zero as the number of party increases (the gap between the local and no-signalling bounds collapses), whereas genuinely multipartite Bell inequalities allow for \emph{non}-decaying witnesses of nonlocality~\cite{WernerWolf,Ineq3PlusChen,MerminForNonlocalFraction}.

Any bipartite scenario can be artificially lifted to a tripartite scenario by adding an extra spectating party~\cite{pironio2005lifting}.
To exclude such uninteresting cases, it is critical to find an appropriate criterion for whether a setup in a tripartite scenario is \emph{genuine}, \emph{i.e.}, exploits possibilities not present in scenarios involving only two parties. 
One avenue to highlight tripartiteness is to focus on entanglement --- the property of quantum states that enables nonlocal correlations. This is the proposal of Ref.~\cite{LOCCInappropriate} which relates nonlocality to the notion of tripartite entanglement formalized in Ref.~\cite{NetworkEntanglement2020}. Such genuinely tripartite entanglement resists any explanation in terms of local operations applied to networks of bipartite quantum states. 

This letter proposes instead a theory-agnostic avenue.
We consider any causal description of Nature --- including classical and quantum physics, and beyond --- and ask the following fundamental question: \emph{Could our physical world be comprised of merely bipartite nonlocal causal constituents?}
That is, does there exist any description of quantum theory's operational predictions, perhaps very exotic, built upon bipartite nonclassical common causes?
It is already well known that bipartite resources are not enough to reproduce all tripartite phenomena.
For instance, perfect correlations between three parties cannot be obtained from bipartite resources, even in a theory-agnostic analysis~\cite{Henson2014}. 
However, that result is predicated on the absence of shared randomness, which is arguably not realistic. Shared classical randomness can be obtained by pre-agreement on a common classical phenomenon to observe, or with preestablished shared randomness stored in local memories.
It is also known that boxworld~\cite{Janotta2012Boxword}, an alternative theory for correlations based on no-signalling boxes~\cite{Barrett2007GPT}, cannot reproduce all quantum correlations even when allowing for shared randomness~\cite{Chao2017genuinemultipartite,Bierhorst2020Tripartite}. This result is restricted to a precise alternative to classical and quantum mechanics, and may not encompass all possible causal theories of correlations~\cite{Chiribella2011Reconstruction, Chiribella_2014}.

Accordingly, in this letter we focus on the (non)simulability of certain tripartite correlations in setups allowing for the local composition of any bipartite resources \emph{with} global access to common shared randomness. 
We adopt a theory-agnostic perspective that applies to any causal theory~\cite{Chiribella2011Reconstruction, Chiribella_2014} compatible with device replication~\cite{PRA} --- however exotic it might be. This includes the classical theory and quantum theory as specific causal theories, but also more generally any hypothetical Generalized Probabilistic Theory (GPT) such as boxword~\cite{Janotta2012Boxword}. Our approach is closely related to the concept of network nonlocality which has been extensively studied in the past decade~\cite{TavakoliReview,fritz2012bell,Branciard2010,Renou2019,Wolfe2019QuantumInflation}.

It is natural to name \emph{genuinely tripartite nonlocal} those correlations which resist explanation in terms of arising from bipartite resources and shared randomness. 
That denotation, however, conflicts with a historical term of art due to Svetlichny~\cite{Svetlichny}. We will explain why Svetlichny's definition is not suitable for causal analysis, leading us to propose an alternative definition (see Definition~\ref{def:GenuineLOSRTripNonloc}), which constitutes the main \emph{conceptual} result of this letter. 

Subsequently, we prove that ${\ket{\rm GHZ}\coloneqq ({\ket{000}+\ket{111}})/{\sqrt{2}}}$ is a resource that can manifest correlations which are genuinely tripartite nonlocal according to our novel definition. This is the subject of Proposition~\ref{prop:GHZ_3}, the main \emph{technical} result of this letter. The formal characterization of such correlations, along with our proof of the quantum realizability of such correlations, together constitute a profound implication: The operational predictions of the quantum theory preclude --- in the strongest possible sense --- any future description of Nature built upon bipartite common causes, regardless of how exotic or nonclassical they could be. 

We conclude this letter by contrasting our no-go theorem with previous works aiming to exclude physical theories limited to 2-way nonclassical common causes. We also recognize the desideratum of certifying Nature's genuine multipartiteness without presupposing the operational validity of quantum theory, and accordingly discuss considerations for the experimental verification of our results.

Although this letter focuses mainly on the tripartite case for pedagogical simplicity, we note that all of our introduced concepts and most of our results are valid in the generalized multipartite case, beyond three parties. We develop the $N$-partite case in an extended version of this work~\cite{PRA}, which includes extending the result regarding the $\ket{\rm GHZ}$ state to any number of parties $N$ (see Proposition~\ref{propo:GHZ_NW_N}) as well as a result regarding the resourcefulness of the ${\ket{\rm W}\coloneqq ({\ket{001}+\ket{010}+\ket{100}})/{\sqrt{3}}}$ state (see Proposition~\ref{propo:W-state}). 
These generalizations of our main results to any number of parties imply that, for any fixed $k$, any theory based on subjecting $k$-way multipartite resources to local operations cannot reproduce the operational predictions of quantum theory for $N{>}k$ spacelike-separated parties.

\begin{figure}
    \centering
        \includegraphics{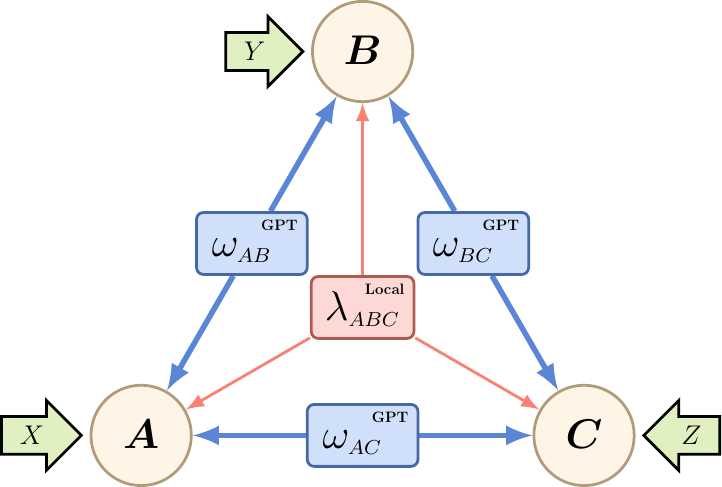}
        
        \caption{A tripartite distribution is \emph{genuinely tripartite nonlocal} according to our definition if it cannot be realized by the above scenario, where the output of each player is determined by local operations (such as joint measurements) on 1)~their input, 2)~the 3-way randomness, and 3)~2-way GPT resources.  \label{fig:GPT_triangle}}

\end{figure}

\emph{A causally meaningful notion of genuine tripartite nonlocality.---}
We seek to distinguish those correlations which admit causal explanation in terms of bipartite nonclassical sources from correlations which resist any such causal explanation. Furthermore, in order to claim that Nature's nonlocality is necessarily tripartite without \emph{a priori} assuming the correctness of quantum causal explanations, we must be careful to apply the label \enquote{genuinely tripartite} only to those correlations which resist bipartite causal explanations in \emph{any} physical theory.

One might ask if Svetlichny's historically accepted \emph{definition} of genuine tripartite nonlocality~\cite{Svetlichny} is suitable for capturing such causal distinction. But no, it is easily hacked: the correlations obtained from CHSH violations in parallel between Alice and Bob as well as between Bob and Charlie fulfill Svetlichny's criterion for genuine tripartite nonlocality~\cite{Tejada2020NetworkGMNL}. Such correlations, however, are facially achievable in 
quantum theory restricted to bipartite states.
What Svetlichny's definition \emph{is} suitable for is as device-independent witness of genuine tripartite entanglement. Note that the traditional definition of genuine tripartite \emph{entanglement} due to~\citet{Seevinck2001} is susceptible to precisely the same sort of hacking: A 4-qubit state composed of a singlet shared between Alice and Bob as well as a singlet shared between Bob and Charlie satisfies Seevinck's criterion for genuine tripartite entanglement, despite factorizing into bipartite constituents.

The reasons why the historical definitions of tripartiteness for both nonlocality and entanglement are ill-suited for causal analysis is because they were motivated by quantifying resourcefulness relative to Local Operations and Classical Communication (LOCC). When analysing Bell-inequality violations, however, we presume that the parties involved may be spacelike separated, which enforces the No-Signalling condition. When classical communication is forbidden, the only form of processing of nonclassical resources that remains is via Local Operations and Shared Randomness (LOSR)~\cite{Wolfe2020quantifyingbell,LOCCInappropriate,sengupta2020quantum}.

Therefore, it is critical to employ the LOSR resource-theoretic framework instead of LOCC when quantifying the nonclassicality of a common cause in a Bell experiment. Ironically, Svetlichny's~\cite{Svetlichny} definition was specifically tailored to the task of witnessing \emph{LOCC} tripartite entanglement, which is irreconcilably in tension with quantifying nonlocality, as nonlocality is only meaningfully studied in the \emph{LOSR} paradigm.

A notion of genuine tripartiteness relative to LOSR entanglement has been formulated in Refs.~\cite{NetworkEntanglement2020,LOCCInappropriate}. Ref.~\cite{LOCCInappropriate} seamlessly extends that notion to provide a definition of genuine tripartite \emph{nonlocality} based on the concept of a correlation resisting explanation in terms of bipartite quantum states acted upon by LOSR. Our main conceptual contribution here is to provide an LOSR-motivated definition for genuine tripartite nonlocality that is theory-agnostic, in that it imagines that LOSR could be applied to \emph{any} sort of bipartite nonclassical resource, not just quantum entanglement.

\begin{figure*}[htb]\centering
    \includegraphics[width=\linewidth]{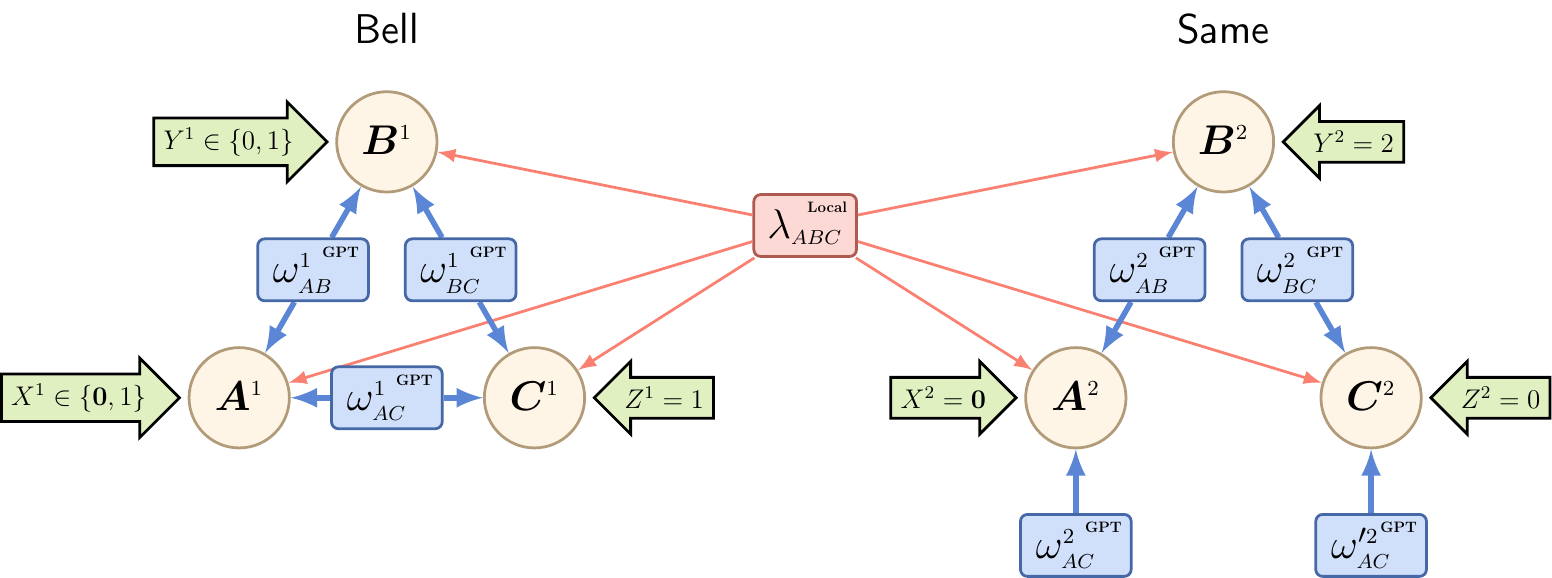}
    \cprotect    \caption{%
    The inflation technique consists of duplicating and rearranging players, sources, and input distributions. 
    Here we inflate the (non genuinely tripartite-nonlocal) triangle scenario of Figure~\ref{fig:GPT_triangle} as to have the players play two parallel games ($\Bell$ and $\same$). It leads to a contradiction with the statistics of measurements on $\ket{\rm GHZ}$, and therefore to the conclusion that the $\ket{\rm GHZ}$ quantum state is a genuinely tripartite-nonlocal resource.
    The duplicated players constitute indistinguishable copies of the same abstract process, hence Alice, on input $X{=}\mathbf{0}$, could be playing either game ($A_1$ and $A_2$ must have the same behaviour). The only condition on the random inputs is that they be independent from all of the sources.
    The figure represents a cut of a larger inflation of order 3, consisting of a triangle and a hexagon.
    \label{fig:MinInflationArgument}}
\end{figure*}\FloatBarrier

We appeal to the GPT formalism to formally define local operations on \enquote{\emph{any} sort of bipartite nonclassical resource.} In brief, we allow for any exotic physical theory that can extend (or restrict) the bipartite resource of quantum entanglement (including all nonsignalling nonlocal boxes such as the PR box~\cite{Popescu1994}), and that can extend (or restrict) the process of combining subsystems via entangled joint quantum measurement~\cite{Branciard2010,BilocalCorrelations}.
Quantum theory itself is merely one of an infinite spectrum of such hypothetical physical theories~\cite{Chiribella2011Reconstruction,Chiribella_2014,Skrzypczyk2009couplers,Short2010couplers,Barrett2007GPT,Janotta2012Boxword}.

\begin{definition}[Genuine LOSR tripartite nonlocality]\label{def:GenuineLOSRTripNonloc}
A tripartite nonsignalling correlation $P$ is said to be \term{genuinely LOSR tripartite nonlocal} if and only if it cannot be obtained by local operations over any 2-way GPT resources along with 3-way shared randomness between all parties. That is, $P$ is said to be genuinely LOSR tripartite nonlocal when it cannot be realized via the abstract causal process depicted in Figure~\ref{fig:GPT_triangle}.
\end{definition}
Equipped with this new definition, let us now provide examples of quantum tripartite resources which are genuinely tripartite nonlocal. 
We also assume that \emph{every} causal theory allow for device replication, i.e. one can make independent and identical copies of resources, to draw inferences from the nonfanout-inflation technique~\cite{Wolfe2016inflation} (see Ref.~\cite{PRA} for an extended formal treatment of these ideas).

\emph{Genuinely tripartite nonlocal correlations exist in Nature.---}
We now prove that ${\ket{\rm GHZ}\coloneqq {\ket{000}+\ket{111}}/{\sqrt{2}}}$ generates quantum correlations which are genuinely LOSR tripartite nonlocal. As in~\cite{Chao2017genuinemultipartite}, the basic idea is to split the problem into two intertwined games, respectively detecting that some party's measurement must depend on both (1)~some nonclassical resource, albeit possibly bipartite, and (2) some tripartite resource, albeit possibly classical. Performing well at both (1)~\emph{and}~(2) would require dependence on a genuinely LOSR tripartite nonclassical (entangled) resource. More precisely, we introduce

\begin{enumerate}
    \item[(1)] A bipartite nonlocal game (conditioned on the third's player output), which rewards nonclassical randomness.
\end{enumerate}    
This first task is the standard CHSH game between Alice and Bob, with the particularity that it is scored only when Charlie outputs $C{=}1$. The function to maximize is (the observables take value in $\{-1,+1\}$)
\begin{align}\label{eq:BellStandard}
I_\Bell^{C_1{=}1}\coloneqq   \langle A_0B_0 &\rangle_{C_1{=}1} + \langle A_0B_1 \rangle_{C_1{=}1} \nonumber\\&+\langle A_1B_0 \rangle_{C_1{=}1} -\langle A_1B_1 \rangle_{C_1{=}1} \,.%
\end{align}

\begin{enumerate} 
\setcounter{enumi}{1}
    \item[(2)] A tripartite consistency game that rewards no-randomness or tripartite randomness.
\end{enumerate}

Here, the players are asked to output the same result (which can take either of the two values~$\pm1$), and are scored according to the function
\begin{align}\label{eq:defsame}
I_\same\coloneqq & \langle A_0B_2 \rangle +\langle B_2C_0 \rangle \,.%
\end{align}

Because $A_0\coloneqq A_{X{=}0}$ belongs to both games, on that input Alice is oblivious as to which of the two games she is partaking in. This prevents her from playing the two games separately; rather, her strategy for ${X=0}$ must be optimized in respect to both games simultaneously. The impossibility of Alice decoupling the two games leads to our central argument:
\begin{equation*}
``\eqref{eq:BellStandard}+ \eqref{eq:defsame} \text{~rewards only genuinely tripartite nonlocality.''}
\end{equation*}

More precisely, in the $\ket{\rm GHZ}$ case, we combine $I_\Bell^{C_1{=}1}$ and $I_\same$ into an inequality:
\begin{prop}[${\rm GHZ_3}$\label{prop:GHZ_3}]
In the absence of any 3-way nonclassical cause, if $\langle C_1 \rangle=0$,
\begin{equation}\label{eq:MainIneqGHZ3}
I_\Bell^{C_1{=}1}+4 I_\same\le 10 \,.
\end{equation}
Measurements on the $\ket{\rm GHZ}$ quantum state can violate the above by reaching $I_\Bell^{C_1{=}1}+4 I_\same=2\sqrt{2}+8>10$. The maximal GPT violation reaches the algebraic maximum of $12$.
\end{prop}

For a better presentation, we focus on explaining why reaching the algebraic maximum of~12 leads to a contradiction. The quantified proof of Ineq.~\eqref{eq:MainIneqGHZ3} is done in~\cite{PRA}, where we also explain how to remove the $\langle C_1 \rangle=0$ assumption (this assumption is experimentally problematic).

\begin{proof}[Proof of equation~\eqref{eq:MainIneqGHZ3}, main ideas.]
Let us assume by contradiction the existence of three black-box devices that satisfy the causal structure of the triangle scenario (Figure~\ref{fig:GPT_triangle}), but that can nevertheless reach the perfect scores ${I_\Bell^{C_1{=}1}=4}$ and ${I_\same=2}$.

Inspired by inflation-technique ideas, we now imagine an inflated scenario where the devices and resources are duplicated and rearranged; see Figure~\ref{fig:MinInflationArgument}. Note that the same instance of the shared randomness $\lambda$ can be infinitely copied and hence be distributed to all parties, but that the (2-way) GPT resources cannot; it is possible, however, to have multiple independent instances of each of those resources by device replication. In our scenario, some of the 2-way resources are inputted only to a single player; their second halves can be considered never measured.

First, on the left-hand side of the figure, the devices take the $\Bell$ test and inherit exactly the behaviour of the original devices (if we ignore the right-hand side of the inflated scenario, the left-hand side is precisely the original scenario).

An important property of Bell inequalities is that any violation implies true randomness~\cite{Pironio2010,AnomalyExtra2018Bamps}. In our case, $A^1B^1$ reaches the maximal algebraic violation of CHSH, which implies that $A^1$ (and also $B^1$) is totally unpredictable. Hence, in particular,
\begin{equation}
 A^1_{X{=}0}C^2_{X{=}0} \text{~are perfectly uncorrelated.}\label{contradiction1}
\end{equation}

Second, on the right-hand side, the devices perform the $\same$ test. As we do not know the inner workings of the black boxes, we cannot describe their whole tripartite joint behaviour. However, note that $A^2B^2$ and $B^2C^2$ inherit the joint statistics of their respective original counterparts, because they see the same environment. 
This means that they achieve perfect correlations at the $\same$ test: ${A^2_{X{=}0}=B^2_{X{=}0}=C^2_{X{=}0}}$. Finally, from the structure of the graph, $A^1C^2$ and $A^2C^2$ also see the same environment and share the same statistics, so
\begin{equation}
 A^1_{X{=}0}C^2_{X{=}0} \text{~are perfectly correlated.}\label{contradiction2}
\end{equation}
    
The contradiction between (\ref{contradiction1})~and~(\ref{contradiction2}) ends our demonstration. In~\cite{PRA} we explain how all the ingredients of this proof can be made quantitative to obtain the trade-off described by Eq.~\eqref{eq:MainIneqGHZ3}.
\end{proof}

\begin{proof}[Proof of violation]
The quantum violation is achieved using $\ket{\rm GHZ}$:
On inputs corresponding to the $\same$ game ($XYZ{=}020$), all players measure in the rectilinear basis. On input $Z{=}1$, Charlie measures his state in the Hadamard basis and obtains marginal $\langle C_1\rangle =0$; when he obtains $C_1{=}1$ (corresponding to a measurement result $\ket{+}_C$), the state of Alice and Bob is steered towards the maximally entangled state $\ket{\phi^+}_{AB}$ and they can play the $\Bell$ game using the standard optimal strategy for CHSH.

Note that the maximal algebraic violation is achieved by the nonsignalling distribution ${A_x{\coloneqq} (-1)^{r_0\oplus r_1 \cdot x}}$, ${B_y{\coloneqq} (-1)^{r_0\oplus x\cdot y}}$, ${C_z{\coloneqq} (-1)^{r_z}}$, where $r_0$ and $r_1$ are uniformly random bits, and $\oplus$ denotes addition modulo $2$.
\end{proof}

\emph{Generalization.---}
In~\cite{PRA}, we show how these ideas can be used to prove a similar result for the ${\ket{\rm W}\coloneqq {\ket{001}+\ket{010}+\ket{100}}/{\sqrt{3}}}$ state.\begin{prop}[${\rm W}$]\label{propo:W-state}
Appropriate measurements on the $\ket{\rm W}$ quantum state lead to genuinely LOSR-tripartite-nonlocal correlations.
\end{prop}
We also explain how to generalize our work to scenarios with arbitrary number of parties, in which $\ket{\rm GHZ}$ straightforwardly generalizes to an $N$-partite state $\ket{{\rm GHZ}_N}$. Indeed, our definition~\ref{def:GenuineLOSRTripNonloc} can be generalized to the multipartite case~\cite{PRA}, introducing the concept of genuine LOSR multipartite nonlocality for which we have:
\begin{prop}[${\rm GHZ}_N$]\label{propo:GHZ_NW_N}
For any $N$, genuinely LOSR $N$-multipartite nonlocal correlations can be obtained through appropriate measurements on the quantum state $\ket{{\rm GHZ}_N}$.
\end{prop}

\emph{Discussion.---}
We have proven that the correlations of $\ket{{\rm GHZ}}$
can only be obtained using genuinely LOSR-tripartite-nonlocal resources. 
Our work implies, under the (reasonable) hypothesis that quantum mechanics' predictions for local measurements over $\ket{{\rm GHZ}}$ are exact, that \emph{Nature cannot be merely bipartite}. 
In~\cite{PRA}, our generalization implies that it cannot even be $N$-partite for any fixed $N$.

In our introduction, we intentionally kept the concept of \emph{combining any exotic GPT bipartite resources}, together with tripartite shared randomness, vague.
Let us now clarify it, based on the nonfanout-inflation technique~\cite[Sec.~5.4]{Wolfe2016inflation}, which is used in our proof (see also other related frameworks~\cite{Henson2014,Chiribella_2014,Chiribella2011Reconstruction,GisinNSI,Bancal2021Networks,Beigi2021,Pironio2021InPreparation}). 
It relies on two postulates. 
First, we admit the possibility of device replication: Any device distributing local resources, or locally operating resources, can be duplicated in independent copies, and one can reorder these replicated devices to form a new setup.
Second, we admit causality. It implies that any two identical subsets of the initial or new setups must have the same behaviour (more than a consequence of causality, this can be seen as an operational definition of what is causality). 
Moreover, for any fixed value of the shared randomness, any marginal correlation of two disjoint subsets of a setup must factorize. 
With inflation, these two postulates provide the definition of \emph{theory-agnostic correlations in networks}, which are all correlations $P$ which do not lead, in any inflated scenario, to any contradiction. See Ref.~\cite{PRA} for the formalized definition.

Let us conclude this letter with experimental considerations. 
In~\cite{PRA}, we relax our experimentally unrealistic constraint $\langle C_1 \rangle=0$ for inequality~\eqref{eq:MainIneqGHZ3} to a generalized inequality valid for all $C_1$. 
Moreover, remark that for a mixture of the $\ket{{\rm GHZ_3}}$ state with white noise, of fidelity $f$, our inequality is violated for $f\gtrsim 93\%$. 
In~\cite{PRA}, we propose an algorithm based on inflation able to witnesses infeasibility down to $f\gtrsim 85\%$. 
This shows that an experimental proof that Nature is not merely bipartite is accessible to current technologies~\cite{Hamel2014GHZ3MerminSvetlichny}. The experimental feasibility for larger $N$ values is an open question~\cite{GHZExperiment6Photons,Chao2019}.

\emph{Acknowledgements.---}
We thank Claude Crépeau, Nicolas Gisin, Miguel Navascués, and Stefano Pironio for valuable discussions.
This research was supported by the Swiss National Science Foundation (SNF), the Fonds de recherche du Qu\'ebec -- Nature et technologies (FRQNT), and Perimeter Institute for Theoretical Physics. Research at Perimeter Institute is supported in part by the Government of Canada through the Department of Innovation, Science and Economic Development Canada and by the Province of Ontario through the Ministry of Colleges and Universities. M.-O.R.~is supported by the Swiss National Fund Early Mobility Grant P2GEP2\_191444 and acknowledges the Government of Spain (FIS2020-TRANQI and Severo Ochoa CEX2019-000910-S), Fundació Cellex, Fundació Mir-Puig, Generalitat de Catalunya (CERCA, AGAUR SGR 1381) and the ERC AdG CERQUTE.

\bigskip
\nocite{apsrev42Control}
\setlength{\bibsep}{1pt plus 1pt minus 1pt}
\bibliographystyle{apsrev4-2-wolfe}
\bibliography{GPTandOtherConstraints}

\end{document}